\documentclass[12pt]{iopart}
\input epsf
\include{psfig}

\newcommand{\BEQ}{\begin{equation}}
\newcommand{\EEQ}{\end{equation}}
\newcommand{\BEA}{\begin{eqnarray}}
\newcommand{\EEA}{\end{eqnarray}}

\renewcommand{\d}{{\rm d }}

\renewcommand{\S}{S_{\rm ep}}
\newcommand{\p}{\partial}

\newcommand{\I}{{\cal I}}

\def\dbarrm {{\mathchar'26\mkern-11mu{\rm d}}}

\newcommand{\fix}{{\Bigl.\Bigr|}}

\begin{document}

\title{Thermodynamic picture of the glassy state}

\author{Th.M. Nieuwenhuizen}
\address{Department of Physics and Astronomy,\\
Valckenierstraat 65, 1018 XE Amsterdam, The Netherlands}

\begin{abstract}

A picture for thermodynamics of the glassy state is introduced.
It assumes that one extra parameter, 
the effective temperature, is needed to
describe the glassy state. This explains the classical paradoxes
concerning the Ehrenfest relations and the Prigogine-Defay ratio.
As a second part, the approach connects the response of
macroscopic observables to a field change with
their temporal fluctuations, and with
the fluctuation-dissipation relation, in a generalized
non-equilibrium way. 

\end{abstract}

\pacs{02.70.Ns, 61.20.Lc, 61.43.Fs}

\submitted{{\noindent \it }}

\section{Introduction}

Thermodynamics for systems far from equilibrium
has long been a field of confusion. A typical application is window glass,
which is far from equilibrium: it 
 is an under-cooled liquid that, in the glass formation process,
 has fallen out of its meta-stable equilibrium. 

Until recently, the general consensus reached
after more than half a century of research was:
{\it Thermodynamics does not work for glasses,
because there is no equilibrium}.%~\cite{Angell}. 
This conclusion was partly based on
the failure to understand the Ehrenfest relations
and the Prigogine-Defay ratio. It should be kept in mind
that, so far, the approaches leaned very much on equilibrium ideas.
Well known works are the 1951 Davies-Jones paper~\cite{DaviesJones},
 the 1958 Gibbs-DiMarzio paper
{}~\cite{GibbsDiMarzio}, % and the 1965 Adam-Gibbs~\cite{AdamGibbs} papers,
while a 1981 paper by DiMarzio has title ``Equilibrium theory of
glasses'' and a subtitle ``An equilibrium theory of glasses is
absolutely necessary''~\cite{DiMarzio1981}.
We shall stress that such approaches are not always applicable, due to the
inherent non-equilibrium character of the glassy state.
Actually, the above conclusion in italics is incorrect itself,
the proper statement being that ``thermostatics does not work for
glasses''. Let us stress that in literature ``thermodynamics of glasses'' 
usually refers to actual systems under laboratory conditions, i.e.
far from equilibrium. These terminologic issues  show how regretful it is
that ``thermostatics''  got lost as expression for 
``equilibrium thermodynamics''.
\indent Thermodynamics is the most robust field of physics.
Its failure to describe the glassy state is  quite unsatisfactory, 
since up to 25  decades in time can be involved.
Naively we expect that each decade has its own dynamics, 
basically independent of the other ones. 
We have found support for this point
in models that can be solved exactly.
Thermodynamics then means a description % of system properties
under smooth non-equilibrium conditions.

%In section 2 we explain the thermodynamic
%picture, and in section 3 we discuss
%models that (partly) support it.

\section{Thermodynamic picture for systems having an
effective temperature} \label{General}

A system that, after a quench to certain low temperature, slowly relaxes 
to equilibrium is characterized by the time elapsed since the 
quench, sometimes called ``age'' or ``waiting time''.
For glassy systems this is of special relevance.
For experiments on spin glasses it is known that non-trivial cooling
or heating trajectories can be described by an effective age~\cite{Hammann}.
Yet we do not wish to discuss spin glasses. They have an
infinity of long time-scales, or infinite order replica symmetry
breaking. 

We shall restrict to systems with one diverging time scale,
having, in the mean field limit, one step of replica symmetry
breaking. They are systems with first-order-type phase transitions,
having discontinuous order parameter, but usually no latent heat.

We consider transitions for glass forming 
liquids as well as for random  magnets.
The results map onto each other by interchanging  volume $V$,
 pressure $p$, compressibility $\kappa=-\p \ln V/\p p$, and
expansivity $\alpha=\p \ln V/\p T$,
 by  magnetization $M$,  field $H$, susceptibility
$\chi=(1/N)\p M/\p H$,  and ``magnetizability'' $\alpha=(-1/N)\p M/\p T$,
respectively.

The picture to be investigated in this work applies to systems
of which the non-equilibrium state involves two well-separated
time-scales. It can then characterized by three parameters,
$T, p$ and the {\it effective temperature} $T_e(t)$. 
In model systems to be discussed below, this quantity follows 
from analytically solving the dynamics of the system;
in realistic (model) glasses that can be approximately described
by a two-timescale picture it could follow from appropriate 
(numerical) experiments ~\cite{Kobby}.
For a set of smoothly related cooling experiments $T_i(t)$
at  pressures $p_i$, one may express the effective
temperature as a continuous function: $T_{e,i}(t)$ $\to$ $T_e(T,p)$.
For the given set of experiments this sets a surface in $(T,T_e,p)$
space. Combining with other experiments, such as cooling at
a different rate, or first cooling, and then heating, the
surface becomes multi-valued.
For covering the whole space one
needs to do many experiments, e.g., at different pressures
and different cooling rates.
The results should agree with findings from heating experiments
and aging experiments.
Thermodynamics amounts to giving differential relations between
observables at nearby points in this space.

Of special interest is the thermodynamics of a thermal body at
temperature $T_2$ in a heat bath at temperature $T_1=T$. 
A basic assumption is separation of time-scales, and consequent
separation of phase space, allowing to identify 
entropies $S_1$ and $S_2$. (For the energy such a decomposition
cannot be made.)
This setup allows to maintain the difference in temperatures, and 
applies to mundane situations such as a cup of coffee, or an ice-cream,
 in a room. 
The change in heat of such
 systems obeys $\dbarrm Q\le T_1\d S_1+T_2\d S_2$.

A similar two-temperature approach proves to be 
relevant for glassy systems.
The known exact results on the thermodynamics of systems can be
% written in the same form, with the equality sign
summarized by the very same change in heat
~\cite{NEhren} \cite{Nthermo}.
 \BEQ \label{dQ=}\dbarrm Q=T\d\S+T_e\d\I\EEQ
where $\S$ is the entropy of the {\bf e}quilibrium
{\bf p}rocesses, i.e. the fast or $\beta$-processes that have
a timescale less than the observation time.
$\I$ is the configurational entropy of the slow or configurational
processes ($\alpha$-processes), also known as
information entropy or complexity. 
In the standard definition~\cite{GibbsDiMarzio} 
the configurational entropy $S_c$ is the entropy
of the glass minus the one of the vibrational modes of the crystal.
For polymers this still includes short-distance rearrangements, which
is a relatively fast mode. It was confirmed numerically
that $S_c$ indeed does not vanish at any temperature,
and it does not fit well to the Adam-Gibbs~\cite{AdamGibbs} 
relation $\tau_{\rm eq}$ $\sim$ $\exp (C/TS_c)$ ~\cite{Binder}.
Our quantity $\I$ only involves long-time processes; the relatively
fast ones are counted in $S_{\rm ep}$. 
To stress that in $\I$ only the slow modes of $S_c$ contribute, it
would deserve a separate name; the most natural one being {\it complexity}.    
The applicability of an Adam-Gibbs-type relation
$\tau_{\rm eq}\sim\exp (C/T\I)$ remains an open issue, but it is 
satisfied in a toy model for the standard folklore of the glassy state
~\cite{LeuzziN}.

If a system has a set of processes $i$ with very different
timescales $\tau_i$ and partial entropies $S_i$, one can define 
$\I$ as the sum of the $S_i$ having $\tau_i>\tau_{\rm eq}/10$, 
where $\tau_{\rm eq}={\rm max}\, \tau_i$ is the
equilibrium relaxation time. 
For a system quenched to a low temperature and
aging there during a time $t$, 
the sum would be restricted to  $\tau_i > t/10$.

\subsection{First and second law}

For a glass forming liquid the first law $\d U=\dbarrm Q+\dbarrm W$
becomes
\BEA \label{thermoglassp}
\label{dUp=}\d U=T\d \S+T_e \d \I-p\d V
\EEA
It is appropriate to define the generalized free enthalpy
\BEA \label{Gp=} G&=&U-T\S-T_e \I+pV\EEA
This is not the standard free enthalpy, since $T_e\neq T$. It satisfies
\BEA
\label{dGp=}\d G&=&-\S\d T-\I\d T_e+V\d p
\EEA

The total entropy is
\BEQ \label{Stot=}S=\S+\I \EEQ
The second law requires $\dbarrm Q\le T\d S$, leading to
%\BEQ 
$(T_e-T)\d\I\le 0$,% \EEQ
which merely says that heat goes from high to low temperatures.
%This has been checked explicitly~\cite{Nlongthermo}.

Since $T_e=T_e(T,p)$, and both entropies are functions of $T$, $T_e$
and $p$, the expression (\ref{dQ=}) yields the specific heat
%\BEA
$C_p={\p Q}/{\p T}\fix_p$.
%T(\frac{\p \S}{\p T}\fix_{T_e,p}+
%\frac{\p \S}{\p T_e}\fix_{T,p}\frac{\p T_e}{\p T}\fix_p)
%+T_e(\frac{\p \I}{\p T}\fix_{T_e,p}+
%\frac{\p \I}{\p T_e}\fix_{T,p}\frac{\p T_e}{\p T}\fix_p)\EEA
In the glass transition region all factors, except $\p_T T_e$,
are basically constant. This leads to
\BEA \label{CpTool}
C_p&=&C_1+C_2\frac{\p T_e}{\p T}\fix_p
\EEA
Precisely this form has been assumed half a century ago
by Tool~\cite{Tool}  as starting point for
the study of caloric behavior in the glass formation region,
and has often been used for the explanation of experiments
{}~\cite{DaviesJones}\cite{Jaeckle86}.
It is  a direct consequence of eq. (\ref{dQ=}).

For magnetic systems the first law brings
\BEA
\label{dU=}\d U&=&T\d \S+T_e \d \I-M\d H
\EEA
As above, one can define the generalized free energy
$F=U-T\S-T_e \I$. It satisfies the relation
$\d F=-\S\d T-\I\d T_e-M\d H $.
In an aging system (fixed $T$ and $H$) the rate of change 
$\dot F=-\I\,\dot T_e$ is usually positive. 
In literature~\cite{AdamGibbs}~\cite{CHS} one has also considered the 
``experimental'' or ``dynamical'' free energy $F_{\rm dyn}=U-T(\S+\I)$.
It evolves as $\dot F_{\rm dyn}=(T_e-T)\dot \I$, thus it is related 
to entropy production, and always negative.

\subsection{Modified Maxwell relation}

For a smooth sequence of cooling procedures of a
glassy liquid, eq. (\ref{dUp=}) implies a modified
Maxwell relation between macroscopic observables
such as $U(t,p)\to U(T,p)= U(T,T_e(T,p),p)$ and $V$.
This solely occurs since $T_e$ is a non-trivial function of 
$T$ and $p$ for the smooth set of experiments under consideration. 

For glass forming liquids it reads
\BEQ \label{modMaxp}
\frac{\p U}{\p p}\fix_T + p\frac{ \p V}{\p p}\fix_T
+T\frac{\p V}{\p T}\fix_p
=T\frac{\p \I}{\p T}\fix_p\,\frac{\p T_e}{\p p}\fix_T-
T\frac{\p \I}{\p p}\fix_T\,\frac{\p T_e}{\p T}\fix_p+
T_e\frac{\p \I}{ \p p}\fix_T
\EEQ
%This is the modified Maxwell relation between observables $U$ and $V$.
In equilibrium $T_e=T$, so the right hand side vanishes.
%, and the standard form is recovered.
For a glassy magnet one has
\BEQ \label{modMaxH}
\frac{\partial U}{\partial H}\fix_T+M-
T\frac{\partial M}{\partial T}\fix_H=
T_e\frac{\partial \I}{\partial H}\fix_T
+T\left(\frac{\partial T_e}{\partial H}\fix_T
\frac{\partial \I}{\partial T}\fix_H
-\frac{\partial T_e}{\partial T}\fix_H
\frac{\partial \I}{\partial H}\fix_T\right)
\EEQ

\subsection{Ehrenfest relations and Prigogine-Defay ratio}

In the glass transition region a glass forming liquid exhibits
smeared jumps in the specific heat $C_p$, the expansivity $\alpha$
and the compressibility $\kappa$. If one forgets about the smearing,
one may consider them as true discontinuities, yielding an analogy
with continuous phase transitions of the classical type.

Denoting the discontinuities as
$\Delta O=O_{\rm liquid}-O_{\rm glass}$, we may
follow Ehrenfest and take the derivative of $\Delta
V(T,p_g(T))=0$. This yields the ``first Ehrenfest relation'' 
\BEQ \label{Ehren1p}
\Delta \alpha=\Delta \kappa \frac{\d p_g}{\d T}\EEQ
while for a glassy magnet
%\BEQ \label{Ehren1H}
$\Delta \alpha=\Delta \chi \,{\d H_g}/{\d T}$.

The conclusion drawn from half a century of research on glass
forming liquids is that this relation is never satisfied
{}~\cite{DaviesJones}\cite{Goldstein}\cite{Jaeckle}~\cite{Angell}.
This has very much hindered progress on a thermodynamical approach.
However, from a theoretical viewpoint it is hard to imagine that
something could go wrong when just taking a derivative.
McKenna~\cite{McKenna}, and, independently, also we~\cite{NEhren},
 have pointed out that this relation is indeed satisfied
automatically, but it is important say what is
meant by $\kappa$ in the glassy state.

Let us make an analogy with spin glasses. In mean field theory
they have infinite order replica symmetry breaking.
  From the early measurements of Canella and Mydosh ~\cite{Mydoshboek}
on AuFe it is known that
the susceptibility depends logarithmically on the frequency, so on
the time scale. The short-time value, called Zero-Field-Cooled (ZFC)
susceptibility is a lower bound, while the long time value, called
Field-Cooled (FC) susceptibility is an upper bound. Let us
use the term ``glassy magnets'' for  spin glasses
with one step of replica symmetry breaking. They are relevant for comparison
with glass forming liquids. For them the situation is worse, as the
ZFC value is discontinuous immediately below $T_g$.
 This explains why already directly below the glass transition
different measurements yield different values for $\kappa$.
These notions are displayed in figure 1.

\begin{figure}[b!] 
\label{chiplot}
\epsfxsize=10cm
\centerline{\epsffile{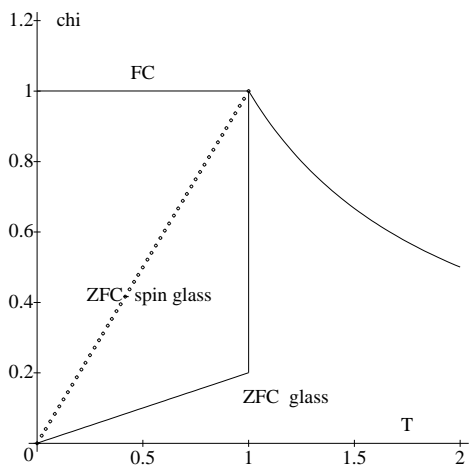}}
\caption{Schematic plot of the field-cooled (FC) and
zero-field-cooled (ZFC) susceptibility in
realistic spin glasses and  in glassy magnets,
as function of temperature, in arbitrary units.
In realistic spin glasses the infinite time or field-cooled
susceptibility is larger than the short time or zero-field-cooled
susceptibility. In magnetic analogs of realistic  glasses
the short time susceptibility even has a smeared  discontinuity at
the glass transition, yielding a value of $\chi$ that depends on 
the precise type of experiment which is performed. In glass forming
liquids the same happens for the compressibility. }
\end{figure}

 Previous claims about the violation of the first Ehrenfest relation
can be traced back to the equilibrium idea that there
 is one, ideal $\kappa$,  to be inserted in (\ref{Ehren1p}).
Indeed, investigators always considered cooling curves $V(T,p_i)$
at a set of pressures $p_i$ to determine $\Delta\alpha$ and
$\d p_g/\d T$. However, $\Delta \kappa$ was always determined in
another way, like measurement of the speed of sound,
or by making pressure steps~\cite{RehageOels}.
In equilibrium such alternative determinations would yield the
same outcome. In glasses this is not the case: the speed of sound is
a short-time process, and additional pressure steps modify the glassy
state.  Therefore alternative procedures should be avoided, and only
the cooling curves  $V(T,p_i)$ should be used. They constitute
a liquid surface $V_{\rm liquid}(T,p)$ and a glass surface
$V_{\rm glass}(T,p)$ in $(T,p,V)$ space. These surfaces intersect,
and the first Ehrenfest relation is no more than a mathematical
identity about the intersection line of these surfaces.
It is therefore automatically satisfied~\cite{NEhren}.
The most careful data we came across were collected by Rehage and
Oels for atactic polystyrene\cite{RehageOels}. In figure 2
we present those data in a 3-d plot, underlining our point of view.
\begin{figure}[b!] 
\label{RehOelsplot3d}
\epsfxsize=15cm
\centerline{\epsffile{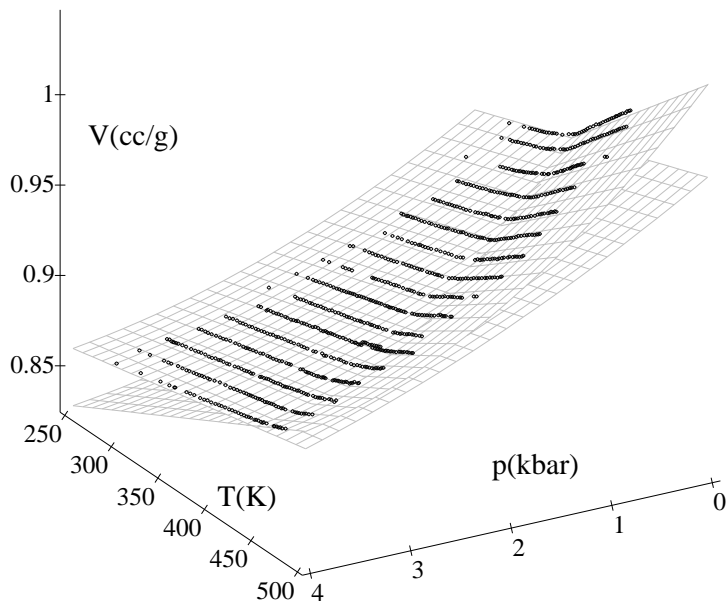}}
\caption{Data of the glass transition for cooling
atactic polystyrene at rate 18 $K/h$, scanned
  from the paper of Rehage and Oels (1976):
specific volume $V$ ($cm^3/g$) versus temperature $T$
($K$) at various pressures $p$ ($k\,bar$). As confirmed by a polynomial
fit, the data in the liquid
essentially lie on a smooth surface, and so do the data in the glass.
The first Ehrenfest relation describes no more than the
intersection of these surfaces, and is therefore
automatically satisfied. The values for the compressibility
derived in this manner will generally differ from results obtained
via other procedures.}
\end{figure}

The second Ehrenfest relation follows from differentiating $\Delta
U(T,p_g(T))=0$. The obtained relation will also be satisfied automatically.
However, one then eliminates $\partial U/\partial p$ by means of
the Maxwell relation. We have already discussed that
outside equilibrium it is modified. The  equality
$T_e(T,p_g(T))=T$ implies 
\BEQ \label{dTedT=1}
\frac{\d T_e}{\d T}=\frac{\p T_e}{\p T}\fix_p+
\frac{\p T_e}{\p p}\fix_T\frac{\d p_g}{\d T}=1 \EEQ
Using eq. (\ref{modMaxp}) and inserting this relation, we obtain
\BEA \label{modEhren2p}
\frac{\Delta C_p}{T_gV}
&=&\Delta\alpha\frac{\d p_g}{\d T}
+\frac{1}{V}\left(1-\frac{\partial T_e}{\partial T}\fix_p\right)
\,\frac{\d \I}{\d T} 
\EEA
The $\d\I/\d T$ term constitutes the total derivative 
along the glass transition line,
$\p \I/\p T+(\p \I/\p p)\d p_g/\d T$.
Its prefactor only vanishes at equilibrium ($T_e=T$), in which case the
standard Ehrenfest relation is recovered.
For glassy magnets one has similarly
\BEQ \label{modEhren2H}
\frac{\Delta C}{NT}=\Delta \alpha\frac{\d H_g}{\d T}+
\frac{1}{N}\left(1-\frac{\partial T_e}{\partial T}\fix_H\right)
\,\frac{\d \I}{\d T}
\EEQ

% Combining the two Ehrenfest relations one may eliminate the
% slope of the transition line. This leads to the so-called
The Prigogine-Defay ratio is defined by
\BEQ
\Pi=\frac{\Delta C_p\Delta\kappa}{TV(\Delta \alpha)^2}
\EEQ
% This looks like an equilibrium quantity.
For equilibrium transitions it should be
equal to unity. Assuming that at the glass transition a number of
unspecified parameters undergo a phase transition, Davies and Jones
derived that $\Pi\ge 1$~\cite{DaviesJones},
 while DiMarzio showed that in that case the correct value is $\Pi=1$
{}~\cite{DiMarzio}.
In glasses typical experimental values are reported in the range
$2<\Pi<5$. It was therefore generally expected that $\Pi\ge 1$ is
a strict inequality arising from the requirement of mechanical stability.

Since the first Ehrenfest relation is satisfied, it holds that
\BEQ\label{Pip=}
\Pi=\frac{\Delta C_p}{T V\Delta\alpha (\d p_g/\d T)}=1+
\frac{1}{V\Delta \alpha }
\left(1-\frac{\partial T_e}{\partial T}\Bigl|_p \Bigr.\right)
\frac {\d \I}{\d p} \EEQ
$\Delta C_p$ and $\Delta\alpha$ can be measured by cooling at
a fixed pressure, but $\d p_g/\d T$ depends on cooling experiments at
two pressures, or, more precisely, on the smooth set of cooling experiments.
 Therefore $\d p_g/\d T$ can be small or large,
implying that $\Pi$ {\it depends on the set of experiments}.
As a result, it can also be below
unity. Rehage-Oels found $\Pi=1.09\approx 1$ at $p=1$
$k\,bar$, using a short-time value for $\kappa$. Reanalyzing their data
we find from (\ref{Pip=}), where the proper long-time $\kappa$ 
has been inserted, a value $\Pi=0.77$. Notice that it is below unity.
The commonly accepted inequality $\Pi\ge 1$ is based on the
equilibrium assumption of a unique $\kappa$. 
Our theoretical arguments and the
Rehage-Oels data show that this assumption is incorrect.

\subsection{Fluctuation formula}

The basic result of statistical physics is that it relates
fluctuations in macroscopic variables to response of their averages
to changes in external field or temperature.
We have wondered whether such relations generalize to the glassy
state. 
% We have found arguments in favor of such a possibility
% both from the fluctuation-dissipation relation and by
% exactly solving the dynamics of model systems ~\cite{Nhammer}.
% Susceptibilities appear to have a non-trivial decomposition, that
% looks as being very general.

In cooling experiments at fixed field it holds that
$M=$$M(T(t),T_e(t,H),H)$. For a thermodynamic description one eliminates
time, implying
$M=$$M(T,T_e(T,H),H)$. One may then expect three terms:
\BEA\label{flucts=}
\chi&\equiv&\frac{1}{N}\,
\frac{\partial M}{\partial H}\Bigl|_T\Bigr.
= \chi^{\rm fluct}(t)+\chi^{\rm loss}(t)+\chi^{\rm conf}(t)
\EEA
The first two terms add up to
%\BEA \label{chifluct1}\chi^{\rm fluct}(t)&=&
$(1/N) ({\partial M}/{\partial H})\Big|_{T,T_e}\Bigr.$.
To find them separately, we switch from a cooling experiment
to an aging experiment at the considered
$T$, $T_e$ and $H$, by keeping, in Gedanken, $T$ fixed from then on.
The system  will continue to age,  expressed by $T_e=T_e(t;T,H)$.
We may then use the equality
\BEQ\label{Mabhelp}
\frac{\partial M}{\partial H}\Big|_{T,t}\Bigr.
=\frac{\partial M}{\partial H}
\Big|_{T,T_e}\Bigr.+ \frac{\partial M}{\partial T_e}
\Big|_{T,H}\Bigr. \frac{\partial T_e}{\partial H}
\Big|_{T,t}\Bigr.
\EEQ
We have conjectured ~\cite{Nhammer}
that the left hand side may be written as the
sum of fluctuation terms for fast and slow processes, defining the
fluctuation contribution
\BEQ\label{Mabhelp1}
\chi^{\rm fluct}(t)=
\frac{\partial M}{\partial H}\Big|_{T,t}\Bigr.
=\frac{ \langle \delta M^2(t)\rangle_{\rm fast}}{NT(t)}+
\frac{\langle \delta M^2(t)\rangle_{\rm slow}}{NT_e(t)}\EEQ
The first term is just the standard equilibrium expression for the fast
equilibrium processes.
Notice that the slow fluctuations enter with their own temperature,
the effective temperature.  This decomposition
is confirmed by use of the fluctuation-dissipation relation
in the form to be discussed below.
From eq. (\ref{Mabhelp}) then follows the ``loss'' term
%and (\ref{Mabhelp1}) yields our non-equilibrium prediction
\BEA \label{chifluct}
\chi^{\rm loss}(t)&=&
%\frac{ \langle \delta M^2(t)\rangle_{\rm fast}}{NT(t)}+
%\frac{\langle \delta M^2(t)\rangle_{\rm slow}}{NT_e(t)}
-\frac{1}{N}\frac{\partial M}{\partial T_e}\Big|_{T,H}\,\,
\frac{\p T_e}{\partial H}\Bigr|_{T,t}
\EEA
%The first two terms are instantaneous, and thus the same for aging
%and cooling. The third term 
It is related to an aging experiment. 
% In the $p$-spin
%model with first $N\to\infty$ and then $t\to\infty$ it vanishes.
In some models it is small~\cite{Nhammer}\cite{Nlongthermo}, but in
another model ~\cite{LeuzziN} it is of order unity.
%Since $T_e\neq T$, t
There occurs in eq. (\ref{flucts=})
also a  configurational term
\BEA \label{chiconf}
\chi^{\rm conf}=\frac{1}{N}\,
\frac{\partial M}{\partial T_e}\Bigl|_{T,H}\,\,
\frac{\partial T_e}{\partial H}\Bigr|_{T}
\EEA
It originates from the difference in  the system's
structure for cooling experiments at nearby fields.
This is the term that is responsible for
 the discontinuity of $\chi$ or $\kappa$ at the glass transition.
Its existence was anticipated by Goldstein and J\"ackle~\cite{Goldstein} 
~\cite{Jaeckle}. 

\subsection{Fluctuation-dissipation relation}

Nowadays quite some attention is payed to the fluctuation-dissipation
relation in the aging regime of glassy systems. It was put forward
in works by Sompolinsky~\cite{Sompolinsky} and 
Horner~\cite{Horner1}\cite{CHS}, and generalized
by Cugliandolo and Kurchan~\cite{CuKu}, 
see ~\cite{BCKM} for a review.

In the aging regime there holds a fluctuation-dissipation 
relation between the correlation function 
$C(t,t')$$=$$<\!\delta M(t)\delta M(t')\!>$ and $G(t,t')$, 
the response of $<\!M(t)\!>$ to a short, small field change 
$\delta H(t')$ applied at an earlier time $t'$,
\BEQ \label{FDR=}
\frac{\partial C(t,t')}{\p t'} = T_e(t,t'){G(t,t')}
\EEQ
with $T_e(t,t')$ being an effective temperature,
also denoted as $T/X(t,t')$~\cite{BCKM}.

We have observed that in simple models without fast
processes $T_e(t,t')=\tilde T_e(t')$
is a function of one of the times only~\cite{Nhammer}~\cite{Nlongthermo}. 
One then expects that $\tilde T_e(t)$ is close to the
``thermodynamic'' effective temperature $T_e(t)$.
We have shown that ~\cite{Nlongthermo}
\BEQ \label{tildeTegen}
\tilde T_e(t)=T_e(t)-\dot T_e(t)
\left(\frac{ \p \ln C(t,t')}
{\p t'}\fix_{t'=t}\right)^{-1}+\cdots
\EEQ
So the effective temperatures $T_e$ and $\tilde T_e$
are not identical. However, in the models analyzed so far,
the difference is subleading in $1/\ln t$.

Notice that the ratio $\p_{t'}C(t,t')/G(t,t')=\tilde T_e(t')$
is allowed to depend on time $t'$.
The situation with constant $T_e$ is well known from mean field spin 
glasses with one step of replica symmetry breaking~\cite{BCKM}, 
but we have not  found such a
constant $T_e$ beyond mean-field~\cite{Nhammer}\cite{Nlongthermo}.
Only at exponential time-scales the mean field spin glass 
behaves as a realistic system~\cite{Nthermo}.

%\newpage

\subsection{Time-scale arguments}

Consider a simple system that has only one type of processes ($\alpha$
processes), which falls out of equilibrium at some low $T$.
When it ages a time $t$ at $T=0$ it will have achieved a state
with effective temperature ${\overline T_e}$, that can be estimated
by equating time with the equilibrium time-scale,
$t=\tau_{eq}({\overline T}_e)$.
We have checked in solvable models that,
to leading order in $\ln t$, it holds that ${\overline T}_e=T_e$.
(The first non-leading order turns out to be  non-universal).
This equality also
is found in cooling trajectories, when the system is well
inside the glassy regime. It says that the
system basically has forgotten its history, and ages on its own,
without caring about the actual temperature.
Another way of saying is that dynamics in each new
time-decade is basically independent of previous decade.

This time-scale argument, however, is not very strong.
Though it works in simple model glasses,
it does not work, for instance, in realistic spin glasses. 

\section{Solvable models}

In the above a variety of effective temperatures have been defined,
and other definitions  appeared also~\cite{CKP}. The most prominent
one seems to us the ``thermodynamical'' $T_e$, that appears in
the second law $\dbarrm Q\le T\d \S+T_e\d \I$.
If these effective temperatures are (basically) the same, 
then the above description
leads to a coherent two-temperature picture. Let us now discuss
models where this is or could be the case.
(In principle also an effective field can occur~\cite{LeuzziN}; 
we shall not consider that complication.)

The thermodynamic part of 
 the above picture has been constructed from aging properties
of the $p$-spin model~\cite{CHS}\cite{NEhren}.
If the limit $N\to\infty$ is taken first, the aging
dynamics starting from a random initial configuration follows from
a set of equations very similar to the
mode-coupling equations of glasses~\cite{CHS}. 
The long-time dynamical properties
also follow from a replica calculation using the marginality 
condition~\cite{KirkpT}. Our first step has been to find the physical 
meaning of the marginal replica free energy. Hereto we analyzed the TAP
partition sum at zero field ~\cite{Ncomplexity} and non-zero field
~\cite{Nthermo}. In both cases we convinced ourselves that the
replica free energy (logarithm of the ordinary partition sum)
indeed coincides with the logarithm of the TAP-partition sum.
In other words, nothing went wrong by using replica's, and 
a physical interpretation should be anticipated.
After the appearance of ref. ~\cite{CKP} we were convinced 
that $T/x$ should be interpreted as thermodynamic effective temperature
~\cite{Nthermo}, and
it was already known that this combination appeared in the
fluctuation-dissipation relation~\cite{CuKu}.
Our studies with Sherrington and Hertz~\cite{HSN} on the
$p$-spin model with a ferromagnetic coupling led to the 
insight that the fluctuation formula (\ref{flucts=}) 
is valid there without the loss term.

Similar results follow for a directed 
polymer model with glassy behavior ~\cite{Ndirpol}\cite{NEhren}.
In this model a directed polymer moves on a flat substrate with randomly
located ridges. It prefers to lie in regions where ridges are widely 
separated.
These Griffiths or Lifshitz singularities are the ``TAP-states'' 
of the problem. They occur often enough
when the transversal width $W$ scales exponentially
in the longitudinal width $W\sim\exp(L^{1/3})$. 
The statics of this model has a Kauzmann transition. 
The long time dynamics involves hopping between these states.
For an ensemble of non-interacting polymers starting from a random
initial configuration, it has been argued that the motion involves
a flow to states with decreasing complexity, as happens
in  the $p$-spin model at exponential timescales~\cite{Nthermo}.
Nevertheless, the dynamics of the model deserves further attention.

Very informative are  models with exactly solvable  
parallel Monte Carlo dynamics,
such as  independent harmonic oscillators~\cite{BPR}~\cite{Nhammer}
\cite{Nlongthermo} or independent spherical spins in a random field
~\cite{Nhammer}\cite{Nlongthermo}. In both cases the equilibrium
timescale follows an Arrhenius law, which leads to glassy
behavior at low temperatures when cooling or when
aging from a random initial configuration. Since these models
are not mean field-like, they have a proper dynamics, which 
satisfies the fluctuation formula (\ref{flucts=}). The effective
temperatures decay as $1/\ln t$, with common prefactor,
but different $1/\ln^2t$ corrections.

A model with a set of fast modes and a set of slow modes,
that have a Vogel-Fulcher-Tammann-Hesse law for the divergence
of the equilibrium time-scale, has also been formulated.
It allows to study glassy dynamics below the Kauzmann
temperature~\cite{LeuzziN}.

The two-temperature approach put forward here also explains 
that thermodynamics applies to black holes~\cite{Nblackhole} and
star clusters~\cite{Nstarcl}.

%\newpage\begin{references}

\section{Acknowledgments}

It is a pleasure to thank A.E. Allahverdyan and L. Leuzzi
for stimulating discussion.

 %references}
 
\end{document}